\documentclass[%
 reprint,
nofootinbib,
 amsmath,amssymb,
 aps,
 prd,
 superscriptaddress
]{revtex4-2}
\usepackage{placeins}
\usepackage{booktabs}
\usepackage{graphicx}% Include figure files
\usepackage{dcolumn}% Align table columns on decimal point
\usepackage{bm}% bold math
\usepackage{multirow}

\usepackage[normalem]{ulem}
\usepackage[usenames,dvipsnames,svgnames,table]{xcolor}

\definecolor{blue_col}{RGB}{0,92,175}
\definecolor{red_col}{RGB}{203,64,66}
\usepackage{hyperref}
\hypersetup{linktocpage,
    colorlinks=true,
    linkcolor=red_col,
    filecolor=Mahogany,      
    urlcolor=blue_col,
    citecolor=blue_col,
}
\begin{document}

% \preprint{APS/123-QED}

\title{Cosmic string gravitational wave backgrounds at LISA:\\II. Reconstruction of conventional signals over astrophysical foregrounds}

\author{Androniki Dimitriou}
\affiliation{Instituto de F\'isica Corpuscular (IFIC), CSIC-Universitat de Val\`{e}ncia, 46980, Val\`{e}ncia, Spain}

\author{Daniel G. Figueroa}
\affiliation{Instituto de F\'isica Corpuscular (IFIC), CSIC-Universitat de Val\`{e}ncia, 46980, Val\`{e}ncia, Spain}

\author{Peera Simakachorn}
\affiliation{Khon Kaen Particle Physics and Cosmology Theory Group (KKPaCT),
Department of Physics, Faculty of Science, Khon Kaen University,
123 Mitraphap Rd., Khon Kaen, 40002, Thailand}

\author{Isak Stomberg}
\affiliation{Instituto de F\'isica Corpuscular (IFIC), CSIC-Universitat de Val\`{e}ncia, 46980, Val\`{e}ncia, Spain}

\author{Bryan Zald\'ivar}
\affiliation{Instituto de F\'isica Corpuscular (IFIC), CSIC-Universitat de Val\`{e}ncia, 46980, Val\`{e}ncia, Spain}

%\emailAdd{androniki.dimitriou@ific.uv.es, %daniel.figueroa@ific.uv.es, %peera.simakachorn@ific.uv.es, %bryan.zaldivar@ific.uv.es}

\date{\today}% It is always \today, today,

\begin{abstract}
We study the reconstruction of conventional cosmic-string signals with LISA in the presence of all major known astrophysical foregrounds expected in the LISA band. These include stellar-origin black-hole binaries (SOBHBs), galactic (WDs) and extragalactic (ExWDs) white dwarfs, extreme-mass-ratio-inspirals (EMRIs), and massive black-hole binaries (MBHBs). Using the Simulation-based Inference package~\href{https://github.com/AndronikiDimitriou/GWBackFinder}{\tt GWBackFinder}, 
we perform a joint inference on the LISA noise, foregrounds, and signal, across a range of injected 
string tensions $G\mu$.
We find that reconstructing tensions with an error $\lesssim 10\%$ requires values as large as $G\mu \gtrsim 10^{-11}$,
{\it i.e.}~a factor $\sim10^5$ larger than previous estimates with no foregrounds, and
$\sim 10^2$ larger compared to estimates accounting only for SOBHB and WD foregrounds. This work is the second in a series initiated in Ref.~\cite{Dimitriou:2025bvq}, which aims to quantify LISA’s ability to measure representative cosmic-string models.
\end{abstract}

\maketitle

%\tableofcontents
\section{\label{sec:intro}Introduction}

A new era of cosmic exploration is underway following the direct detection of gravitational waves (GWs) 
by the LIGO/Virgo/KAGRA (LVK) collaboration in the $\sim 10-100$ {\tt Hz} band~\cite{LIGOScientific:2016aoc,LIGOScientific:2017vwq,LIGOScientific:2018mvr,LIGOScientific:2020ibl,LIGOScientific:2021djp,LIGOScientific:2021psn,LIGOScientific:2026wfs,LIGOScientific:2026ctl}, together with compelling evidence reported by pulsar timing array (PTA) collaborations for a GW background (GWB) around $\sim$ {\tt nHz} frequencies~\cite{NANOGrav:2023gor,Antoniadis:2023ott,Reardon:2023gzh,Xu:2023wog}. In the coming decade, next-generation GW observatories, including ground-based detectors such as the Einstein Telescope (ET)~\cite{Hild:2010id,Punturo:2010zz,Abac:2025saz} and Cosmic Explorer (CE)~\cite{LIGOScientific:2016wof,Reitze:2019iox}, as well as the space-based Laser Interferometer Space Antenna (LISA)~\cite{LISA:2017pwj,LISACosmologyWorkingGroup:2022jok,LISA:2024hlh}, will greatly expand GW sensitivity across the frequency spectrum. In this work we focus on LISA, which 
will be probing the $\sim$ {\tt mHz} region, between PTAs and ground-based detectors.

While a large number of compact binaries are expected to emit GWs in the LISA band, many of these will remain individually unresolved. As a consequence, their cumulative emission will generate stochastic signals--- one per binary population---, that will be perceived as foregrounds in the experiment~\cite{Korol:2021pun,Babak:2023lro,Pozzoli:2023kxy,Staelens:2023xjn,Hofman:2024xar,Perego:2025bif}. In addition, a variety of cosmological GWBs may permeate the Universe~\cite{Caprini:2018mtu}, some of which might be detectable by LISA. Such backgrounds can be sourced by a variety of early Universe mechanisms, including inflation~\cite{Grishchuk:1974ny,Starobinsky:1979ty,Rubakov:1982df,Fabbri:1983us,Anber:2006xt,Sorbo:2011rz,Pajer:2013fsa,Adshead:2013qp,Adshead:2013nka,Maleknejad:2016qjz,Dimastrogiovanni:2016fuu,Namba:2015gja,Ferreira:2015omg,Peloso:2016gqs,Domcke:2016bkh,Caldwell:2017chz,Guzzetti:2016mkm,Bartolo:2016ami,Fumagalli:2020nvq,Fumagalli:2021mpc}, post-inflationary particle production~\cite{Easther:2006gt,GarciaBellido:2007dg,GarciaBellido:2007af,Dufaux:2007pt,Dufaux:2008dn,Dufaux:2010cf,Bethke:2013aba,Bethke:2013vca,Enqvist:2012tc,Figueroa:2017vfa,Adshead:2018doq,Adshead:2019lbr,Adshead:2019igv}, kination~\cite{Giovannini:1998bp,Giovannini:1999bh,Boyle:2007zx,Li:2016mmc,Li:2021htg,Figueroa:2018twl,Figueroa:2019paj,Gouttenoire:2021wzu,Co:2021lkc,Gouttenoire:2021jhk,Oikonomou:2023qfz,Eroncel:2025bcb}, thermal plasma~\cite{Ghiglieri:2015nfa,Ghiglieri:2020mhm,Ringwald:2020ist,Ringwald:2022xif,Ghiglieri:2022rfp,Ghiglieri:2024ghm}, oscillons~\cite{Zhou:2013tsa,Antusch:2016con,Antusch:2017vga,Liu:2017hua,Amin:2018xfe}, first-order phase transitions~\cite{Kamionkowski:1993fg,Caprini:2007xq,Huber:2008hg,Hindmarsh:2013xza,Hindmarsh:2015qta,Caprini:2015zlo,Hindmarsh:2017gnf,Cutting:2018tjt,Cutting:2019zws,Pol:2019yex,Caprini:2019egz,Cutting:2020nla,Han:2023olf,Ashoorioon:2022raz,Athron:2023mer,Li:2023yaj,Jinno:2022mie,Caprini:2024gyk}, cosmic defects~\cite{Vachaspati:1984gt,Sakellariadou:1990ne,Damour:2000wa,Damour:2001bk,Damour:2004kw,Fenu:2009qf,Figueroa:2012kw,Hiramatsu:2013qaa,Blanco-Pillado:2017oxo,Auclair:2019wcv,Chang:2019mza,Gouttenoire:2019kij,Chang:2021afa,Figueroa:2020lvo,Gorghetto:2021fsn,Yamada:2022aax,Yamada:2022imq,Kitajima:2023cek,Servant:2023tua}, and large scalar fluctuations~\cite{Matarrese:1992rp,Matarrese:1993zf,Matarrese:1997ay,Nakamura:2004rm,Ananda:2006af,Baumann:2007zm,Domenech:2021ztg,Dandoy:2023jot}. See~\cite{Caprini:2018mtu} for a comprehensive review.

The Universe is therefore likely to be filled with a variety of astrophysical and cosmological GWBs. For example, while the PTA signal is likely due to supermassive black hole binaries (SMBHBs)~\cite{Kelley:2017lek,NANOGrav:2023hfp,Antoniadis:2023xlr}, cosmological backgrounds also represent a viable explanation~\cite{NANOGrav:2023hvm,Antoniadis:2023xlr,Figueroa:2023zhu}. The detection of a GWB from the early Universe will open a new window into beyond-the-Standard-Model (BSM) physics, at scales far above the reach of particle colliders. In this work, we focus on the ability of LISA to probe the GWB generated by \emph{cosmic strings}~\cite{Kibble:1976sj}. 

Cosmic strings are one-dimensional topological defects, naturally expected in many BSM scenarios~\cite{Kibble:1976sj,Hindmarsh:1994re,Vilenkin:2000jqa,Jeannerot:2003qv}. Once created, they form a {\it network} consisting of {\it long} strings that stretch over the cosmological horizon, and a population of sub-horizon closed strings--- {\it loops} ---. After production, a string network rapidly evolves towards a {\it scaling regime}, where the energy density of the string network tracks the total energy density of the universe, allowing them to remain subdominant and cosmologically safe throughout cosmic history~\cite{Vilenkin:2000jqa}. The scaling behavior is realized because the string network can lose energy through various mechanisms, which depend on the underlying nature of the strings. For example, local and global cosmic strings exhibit very different behaviors. In the Nambu–Goto (NG) approximation, where local strings are treated as infinitely thin, loops lose their energy predominantly through GW emission, giving rise to a potentially observable GWB~\cite{Vilenkin:1981bx,Hogan:1984is,Vachaspati:1984gt,Damour:2004kw,Sousa:2013aaa,Blanco-Pillado:2017rnf,Blanco-Pillado:2017oxo,Auclair:2019wcv,Gouttenoire:2019kij,Boileau:2021gbr,Yamada:2022aax,Yamada:2022imq,Servant:2023tua,Blanco-Pillado:2024aca,Wachter:2024zly,Schmitz:2024gds,Avgoustidis:2025svu}. In contrast, global strings lose their energy more efficiently through massless Goldstone emission~\cite{Saurabh:2020pqe,Baeza-Ballesteros:2023say}, so their GWB~\cite{Chang:2019mza,Gouttenoire:2019kij,Chang:2021afa,Gorghetto:2021fsn,Servant:2023mwt} is suppressed compared to the local strings of the same energy scale. Moreover, beyond the NG limit, field-theory simulations find that even local strings can substantially lose their energy through particle production~\cite{Figueroa:2012kw,Hindmarsh:2017qff,Matsunami:2019fss,Auclair:2019jip,Saurabh:2020pqe,Figueroa:2020lvo,Hindmarsh:2021mnl,Baeza-Ballesteros:2023say,Baeza-Ballesteros:2024otj}, implying a significant suppression of the expected GWB~\cite{Hindmarsh:2022awe,Kume:2024adn,Baeza-Ballesteros:2024otj}.

A catalog of GWB signal templates from cosmic-string networks, based on relevant models from the literature, was presented recently in Ref.~\cite{Dimitriou:2025bvq}, from now on referred to as {\tt Paper\,\,I}. Templates were classified as {\it conventional}, based on the NG limit and standard $\Lambda$CDM cosmology, and {\it beyond conventional}, based on modifications of either the loop number density, the expansion history, or the loop properties. Using Simulation-Based Inference (SBI) data-analysis techniques~\cite{Dimitriou:2023knw,Alvey:2023npw,Srinivasan:2025etu}, {\tt Paper\,\,I} initiated a broad program for quantifying LISA's ability to measure GWBs from representative cosmic string models: it quantified the reconstruction precision of the signals in a case-by-case basis, provided handy figures-of-merit of each scenario, and determined the parameter space accessible to LISA for each modeling, including model comparisons for a few examples. Such endeavor, however broad in scope, was only carried out in the absence of astrophysical foregrounds. In the present paper, we amend this aspect, addressing the detectability of cosmic-string signals in the presence of all relevant astrophysical foregrounds expected in the LISA window.

The impact of foregrounds on template reconstruction has been considered before, {\it e.g.}~Refs.~\cite{Baghi:2023qnq,Muratore:2023gxh,Kume:2024xvh,Caprini:2024hue,LISACosmologyWorkingGroup:2024hsc,LISACosmologyWorkingGroup:2025vdz}, and in particular Refs.~\cite{Boileau:2021gbr,Blanco-Pillado:2024aca} on the reconstruction of cosmic-string templates. In the present work we study the detectability of cosmic-string signals over all relevant astrophysical foregrounds known within the LISA window. Namely, we consider the foregrounds from stellar-origin black-hole binaries (SOBHBs)~\cite{Pieroni:2020rob,Babak:2023lro,Lehoucq:2023zlt}, galactic white dwarf binaries (WDs)~\cite{Korol:2021pun}, extragalactic white dwarf binaries (ExWDs)~\cite{Staelens:2023xjn,Hofman:2024xar}, extreme-mass-ratio-inspirals (EMRIs)~\cite{Korol:2021pun,Pozzoli:2023kxy}, and massive black-hole binaries (MBHBs)~\cite{Sesana:2008mz,Perego:2025bif}. While  the latter contributes rather marginally, the superposition of the former signals signals 
contribute to a very large foreground in the LISA band. In this paper, we quantify the impact of all the above foregrounds on the detectability of conventional cosmic-string backgrounds. We provide in this way, the first realistic assessment on the parameter space detectable by LISA of cosmic-string signals in the presence of all leading foregrounds expected in LISA. 

Since our goal is to demonstrate the effect of astrophysical foregrounds on the reconstruction of cosmic-string signals, we focus on the NG limit of local strings, which is the most common case considered in the literature. Furthermore, this case maximizes the signal amplitude (for a given energy scale), and hence the chances of detection by LISA. As we show in our results, even for a signal as large as the conventional cosmic-string GWB considered here, the reconstruction of the background suffers a significant degradation due to the presence of the foregrounds. Our results are in fact quite different--- more pessimistic --- from those presented in Refs.~\cite{Boileau:2021gbr,Blanco-Pillado:2024aca}, where only SOBHBs and WDs were added in the foreground budget, but not ExWDs or EMRIs (neither MBHBs).

This paper is divided as follows. Section~\ref{sec:Backgrounds} describes the conventional cosmic-string GWB model that we use in our analysis, and surveys the primary astrophysical foregrounds relevant to LISA. In Sect.~\ref{sec:results} we present our reconstruction results based on SBI, including comparisons to the foreground-free and reduced-foreground cases. Different reconstruction methods, namely Monte Carlo Markov Chain (MCMC) and Fisher analysis, are also tested against our SBI results. In Sect.~\ref{Conclusions} we present our conclusions. The Appendices provide further technical details on additional reconstruction cases (App.~\ref{app:more_fid}), the LISA noise model and data generation (App.~\ref{app:lisa_noise}), our statistical approach (App.~\ref{app:SBI}), the likelihood used for MCMC (App.~\ref{app:mcmc_likelihood}), and SBI consistency checks (App.~\ref{app:coverage}).

\section{Gravitational Wave Backgrounds}
\label{sec:Backgrounds}

As the major goal of this paper is to demonstrate the effect of astrophysical foregrounds on the reconstruction of cosmic-string signals by LISA, here we introduce first the modeling we chose for the cosmic-string signal (Sect.~\ref{sec:signal}), and then discuss the astrophysical foregrounds expected within the LISA frequency window (Sect.~\ref{sec:foregrounds}).

\subsection{GWB from Cosmic Strings}
\label{sec:signal}

Cosmic strings can be characterized by their energy per unit length $E/L$. For wiggle-less strings--- as those considered here ---, this is given by the string tension $\mu$, which relates the energy scale $\eta$ of the cosmological phase transition that formed the strings. A string tension is typically written in a dimensionless form,
\begin{align}
	G\mu = \left(\frac{\eta}{M_{\rm Pl}}\right)^2 \simeq 2.7 \times 10^{-10} \left(\frac{\eta}{2 \times 10^{15} \, \rm GeV}\right)^2 \, ,
\end{align}
where $G = M_{\rm Pl}^{-2}$, with $M_{\rm Pl} \approx 1.22 \times 10^{19} \, {\rm GeV}$ the (full) Planck mass. Cosmic strings with larger tensions are more massive, implying that their motion can generate a stronger GW signal. Detailed computation of the cosmic-string GWB spectrum exhibits a non-trivial dependence of the amplitude and shape on $G\mu$. As discussed in {\tt Paper\,\,I}~\cite{Dimitriou:2025bvq}, the calculation of the GWB spectrum from cosmic strings %in various setups 
simply breaks down into evaluating three ingredients: \textit{i)} the GW emission power from each loop, 
\textit{ii)} the loop number density of the string network, and \textit{iii)} the cosmic history along which the %produced 
GWs propagate,
\begin{align}
    \Omega_{\rm GW}(f) &= \frac{1}{3 H_0^2 m_{\rm Pl}^2}\sum_{j=1}^{\infty} \underbrace{\frac{2j}{f} ( G \mu^2 P_j )}_{i)} \nonumber\\
    ~ ~ \times&\int_{a(t_{\rm min})}^{a_0} da \, \underbrace{\frac{1}{H(a)}\left(\frac{a}{a_0}\right)^4}_{iii)} \, \underbrace{{\tt n}\left[\frac{2j}{f} \cdot \frac{a}{a_0},t(a)\right]}_{ii)}~,
    \label{eq:master_formula_GWB_strings}
\end{align}
where $m_{\rm Pl} = M_{\rm Pl}/\sqrt{8\pi}$ is the reduced Planck mass. 

In this work, we consider GWB signals from the NG-limit of local strings, and assume a standard $\Lambda$ Cold Dark Matter ($\Lambda {\rm CDM}$) evolution of the universe. The three ingredients required to calculate the GWB signal, are:

\textit{i) GW emission power from one loop.---}When a NG string loop of length $l$ oscillates, it emits GWs with energy power $\mathcal{P}_{\rm GW} = \sum_j P_j G \mu^2$, where the summation accounts for the total emission from the different harmonics of the loop oscillation. The spectrum of its emission power $P_j$ per $j^{\rm th}$-mode, is obtained from numerical simulations of NG strings~\cite{Blanco-Pillado:2017oxo},  
with normalization condition $\Gamma \equiv \sum_j P_j \simeq 51.43$. By losing energy via GWs, a loop of initial length $l_i \sim \mathcal{O}(0.1\,t_i)$~\cite{Blanco-Pillado:2013qja}, shrinks over time and has its length at any time $t$ as $l(t) = l_i - \Gamma G\mu(t-t_i)$ where $t_i$ is the loop formation time. The GW emitted by this oscillating and shrinking loop, has a frequency today determined by the redshifting of the its length scale at the time of emission $t_e$, $f = [2j/l(t_e)] \times [a(t_e)/a_0]$, where the subscript 0 denotes today. The scale factor $a(t)$ of the cosmic expansion can be obtained from solving the Friedmann equation \eqref{eq:friedmann_LambdaCDM} below. We note that recent NG string simulations show that a string loop loses its energy at varying rate over time due to back-reaction effects~\cite{Wachter:2024aos}, leading to a final GWB amplitude %($\Omega_{\rm GW}$) 
that is about 20\% smaller~\cite{Wachter:2024zly} than what we use in this work.

\textit{ii) Loop number density.}---To account for all loops of length $l$ at time $t$, we calculate the  
loop number-density ${\tt n}(l,t) \equiv \int_{t_{\rm min}}^{t} {\tt f}[l_i,t_i] \left[{a(t_i)}/{a(t)}\right]^3 dt_i$, where $t_{\rm min}$ is the time when the network achieved scaling. The loop production function ${\tt f}[l_i,t_i]$ is measured directly from the simulations of the NG string network evolving in radiation and matter domination eras \cite{Blanco-Pillado:2013qja},
\begin{align}
    {\tt f}[l_i,t_i] = \begin{cases}
    {\rm (radiation)} ~ \frac{92.126}{d_H^{5}(t_i)} \,\delta\left[\frac{l_i}{d_H(t_i)}- 0.05\right] ,\vspace{0.3cm}\\ {\rm (matter)} ~ 
    \frac{5.34}{d_H^{5}(t_i)}\left[\frac{d_H(t_i)}{l_i}\right]^{1.69} \\[0.75em]
    ~ ~  \times  \Theta\left[0.06 - \frac{l_i}{d_H(t_i)} \right] \Theta\left[\frac{l_i}{d_H(t_i)} - \Gamma G \mu \right],
    \end{cases}
    \label{eq:loop_production_function_bos}
\end{align}
	with $d_H(t) = a(t)\int_{0}^t a^{-1}(t')dt'$ the horizon distance at cosmic time $t$.

\textit{iii) Cosmic history.}---We consider the $\Lambda {\rm CDM}$ universe where the rate of the cosmic expansion is described by the Friedmann equation through the Hubble parameter, $H(t) = {d\log a}/{dt}$, which is obtained from
    \begin{align}
		\frac{H(t)}{H_0} = \sqrt{\Omega_{\rm rad}^{(0)} \, \mathcal{G}[T(t)] \left[\frac{a_0}{a(t)}\right]^4 + \Omega_{\rm mat}^{(0)}\left[\frac{a_0}{a(t)}\right]^3 + \Omega_{\rm de}^{(0)}},
        \label{eq:friedmann_LambdaCDM}
	\end{align}
where $H_0 = 100\,h \, {\rm km \, s^{-1} \, Mpc^{-1}}$, with $h = 0.6789$ the Hubble parameter today, and the components of energy budget of the universe today are $\Omega_{\rm rad}^{(0)} = 9.2 \times 10^{-5}$ (radiation), $\Omega_{\rm mat}^{(0)} = 0.308$ (matter), and $\Omega_{\rm de}^{(0)} = 0.692$ (dark energy) \cite{Planck:2018vyg}.
The changes of relativistic degrees of freedom participating in the energy density and the entropy density of the Standard Model thermal plasma is accounted by the function $\mathcal{G}(T) = [g_*(T)/g_*(T_0)][g_{*s}(T_0)/g_{*s}(T)]^{4/3}$; we took the $g_*, g_{*s}$ results from App.~C of \cite{Saikawa:2018rcs}, where the effect of interactions among particles is included.

By fixing the above three ingredients, one obtains the NG local-string GWB template that depends only on $G\mu$, known as the ``BOS" modeling~\cite{Blanco-Pillado:2013qja} (see Sect.~3.2 of {\tt Paper\,\,I}~\cite{Dimitriou:2025bvq} for an extensive discussion on it). Tabulated values of $\{f,h^2\Omega_{\rm GW}\}$ can be found at this \href{https://github.com/peerasima/cosmic-strings-GWB/blob/main/templates/standard_local_VOS_BOS/BOS_saikawa.dat}{repository}. The latest constraint that PTA puts upon this template is $G\mu \lesssim 1.9 \times 10^{-10}$ \cite{NANOGrav:2023hvm}.
As shown in {\tt Paper\,\,I}, in the absence of astrophysical foregrounds, LISA can reconstruct this template with an uncertainty [defined in terms of  Eq.~\eqref{eq:precision}] in $G\mu$ smaller than $5\%$, $20\%$, and $30\%$, for signals with $G\mu > 10^{-13}, \, 10^{-15}$ and $10^{-16}$, respectively, 

We note that there exists another conventional template commonly used in the literature, the ``VOS" modeling~\cite{Martins:2000cs}. 
The main difference from the BOS model adopted here, is that the loop number density in the VOS template does not rely on the numerical simulations, but on semi-analytical calculations. Given that, for $G\mu > 10^{-18}$, the difference between the two conventional templates is $\lesssim 20\%$ in $\Omega_{\rm GW}$ amplitude (at all LISA frequencies), in order to distinguish between the two templates {\tt Paper\,\,I} showed that LISA is required to reconstruct the GWB with  a precision $\lesssim 3\%$ in the absence of foregrounds. Since, as we shall show later, astrophysical foregrounds will introduce a much larger reconstruction uncertainty, using the VOS template would essentially lead to the same results as using the BOS modeling, barring the precise numerical values.

\subsection{Astrophysical Foregrounds}
\label{sec:foregrounds}

Astrophysical compact binaries expected to emit GWs in the LISA band, range from massive black hole binaries (MBHBs), to extreme mass-ratio inspirals (EMRIs), extragalactic sources such as 
stellar-origin black hole binaries (SOBHBs), and compact binaries such as white dwarfs (WDs) from our galaxy, or extragalactic white dwarf pairs (ExWDs). LISA, however, will not be able to resolve many of those individual sources. As a result, their cumulative signals will pile up creating a large foreground in the detector, contributed by the many binary population species. For LISA, self-consistent computation of the unresolved GWBs produced by the sum of above mentioned astrophysical populations, has been computed using the iterative subtraction method from Ref.~\cite{Karnesis:2021tsh}, for individual populations~\cite{Korol:2021pun,Babak:2023lro,Staelens:2023xjn,Pozzoli:2023kxy,Hofman:2024xar}, and for all said populations considered simultaneously~\cite{Perego:2025bif}.  

The presence of foregrounds in LISA will challenge the detection capabilities of cosmological GWBs, as both will appear as stochastic signals in the LISA data stream. In this work, we consider the leading astrophysical foregrounds expected in LISA, including contributions from MBHBs~\cite{Sesana:2008mz,Perego:2025bif}, EMRIs~\cite{Korol:2021pun,Pozzoli:2023kxy}, SOBHBs~\cite{Babak:2023lro,Lehoucq:2023zlt}, WDs~\cite{Korol:2021pun} and ExWDs~\cite{Staelens:2023xjn,Hofman:2024xar}. We consider the state-of-the-art computation of the spectral shape of these foregrounds within the LISA band, while their amplitudes are varied over ranges  
due to  
uncertainty in the population modelings explored in the above references. 

\textit{a) Galactic binaries.}---The galactic foreground from unresolved white dwarf binaries (WDs) within the Milky
Way, is modeled as~\cite{Karnesis:2021tsh}
\begin{align}
&h^2\Omega_{\rm WD}(f) = \nonumber\\
&~
\frac{A_{\rm WD}}{2}
\left(\frac{f}{1 \, \mathrm{Hz}}\right)^{2/3}
\left[
1+\tanh\!\left(\frac{f_{k}-f}{f_2}\right)
\right] e^{-(f/f_1)^\nu},
\end{align}
where the frequencies $f_1$, $f_{k}$, and $f_2$ are defined by
\begin{align}
\log_{10}\!\left(\frac{f_1}{\mathrm{Hz}}\right)
&=
a_1\log_{10}\!\left(\frac{T_{\rm obs}}{\mathrm{yr}}\right)+b_1,
\\
\log_{10}\!\left(\frac{f_{k}}{\mathrm{Hz}}\right)
&=
a_k\log_{10}\!\left(\frac{T_{\rm obs}}{\mathrm{yr}}\right)+b_k \, ,
\end{align}
and $f_2=6.7\times10^{-4}$ Hz, respectively, with $a_1=-0.15$, $b_1=-2.72$, $a_k=-0.37$,
$b_k=-2.49$, and $\nu=1.56$.
We model this foreground with a Gaussian prior in $\log_{10} A_{\rm Gal}$ whose mean value is $-7.84$ \cite{Karnesis:2021tsh}
and the standard deviation is $\sigma=0.21$ \cite{LISACosmologyWorkingGroup:2025vdz}.

\textit{b) Stellar-origin black-hole binaries.}---The GW foreground from unresolved stellar-origin black-hole binaries (SOBHBs), based on the population model based on observations at LIGO-Virgo-KAGRA \cite{KAGRA:2021duu,KAGRA:2021kbb}, is modeled as an isotropic GWB, 
\begin{align}
h^2\Omega_{\rm SOBHB}(f)
=
A_{\rm SOBHB}
\left(\frac{f}{10^{-3}\,\mathrm{Hz}}\right)^{2/3}.
\end{align}
We assume a Gaussian prior on $\log_{10}A_{\rm ExB}$ with mean $-12.38$ \cite{Babak:2023lro}
and standard deviation $\sigma=0.344$ \cite{LISACosmologyWorkingGroup:2025vdz}.

\textit{c) Extragalactic white-dwarf binaries.}---The foreground from unresolved extragalactic white dwarf binaries (ExWDs) is modeled as \cite{Staelens:2023xjn, Hofman:2024xar}
\begin{align}
&h^2\Omega_{\rm ExWD}(f)
=\nonumber\\
& ~ ~ A_{\rm ExWD}
\left(\frac{f}{\hat f}\right)^{0.741}
\left[
1+\left(\frac{f}{\hat f}\right)^{4.15}
\right]^{-0.255}
e^{-Bf^3},
\label{eq:WD-foreground}
\end{align}
where $B=1.54\times10^4 ~ {\rm Hz}^{-3}$ and
$\hat f=7.2\times10^{-3} ~ {\rm Hz}$. We assume that this foreground has a Gaussian prior in
$\log_{10}A_{\rm ExWD}$ with mean $-11.06$ and the standard deviation
$\sigma=0.17$, which are estimated to cover the uncertainty band arising in the model of \cite{Hofman:2024xar} (i.e., light green band in their Fig.~8).
Note that a more recent estimate \cite{Boileau:2025jkv}, using different modelings of the binaries' interaction, agrees well with \cite{Staelens:2023xjn, Hofman:2024xar}, except a sharper spectral cutoff around $\sim7~{\rm mHz}$.
In this work, we strict to the result of \cite{Staelens:2023xjn, Hofman:2024xar} where the effect of astrophysical foreground is more prominent.

\textit{d) Extreme mass-ratio inspirals.}---Also relevant for LISA's window,
there is a foreground contribution expected from extreme mass ratio
inspirals (EMRIs). We do not use an analytical template for it, but rather
tabulated data obtained from Ref.~\cite{Perego:2025bif}. The EMRI
contribution is written as
\begin{align}
h^2\Omega_{\rm EMRI}(f)
=
A_{\rm EMRI}\,
S_{\rm EMRI}^{\rm tab}(f),
\end{align}
where $S_{\rm EMRI}^{\rm tab}(f)$ denotes the tabulated spectral shape,
normalized to unity at the reference frequency $f = 3\,{\rm mHz}$, so that
$A_{\rm EMRI} = h^2\Omega_{\rm EMRI}(f = 3\,{\rm mHz})$. The fiducial
tabulated spectrum of Ref.~\cite{Perego:2025bif} corresponds to
$A_{\rm EMRI} \simeq 4.68\times10^{-12}$, i.e.\
$\log_{10}A_{\rm EMRI} = -11.34$. In our analysis, the amplitude $A_{\rm EMRI}$ is treated
as a free parameter and jointly inferred with the other foreground and
noise parameters; to account for possible uncertainties in the foreground
modeling, we assume a Gaussian prior on $\log_{10}A_{\rm EMRI}$ centered
on the above fiducial value, with standard deviation $\sigma = 0.5$.

\textit{e) Massive black-hole binaries.}---We additionally include the foreground contribution
from massive black hole binaries (MBHBs), also taken from the tabulated data of Ref.~\cite{Perego:2025bif} from their HS scenario, 
\begin{align}
\label{eq:mbhb_template}
h^2\Omega_{\rm MBHB}(f)
= A_{\rm MBHB}
S^{\rm tab}_{\rm MBHB}(f).
\end{align}
Here $S^{\rm tab}_{\rm MBHB}(f)$ denotes the tabulated spectral shape,
normalized to unity at the reference frequency $f = 10^{-4}\,{\rm Hz}$, so
that $A_{\rm MBHB} = h^2\Omega_{\rm MBHB}(f = 10^{-4}\,{\rm Hz})$; the
fiducial tabulated spectrum corresponds to
$A_{\rm MBHB} \simeq 1.07\times10^{-12}$, i.e.\
$\log_{10}A_{\rm MBHB} = -11.97$.

\begin{table}[t]
    \begin{tabular}{|c|c|c|}
    \hline
    ~ Parameter ~ & ~ Fiducial value ~ & ~ $\sigma$ (Gaussian prior) ~ \\
    \hline
     $\log_{10} A_{\rm WD}$    & $-7.84$  & 0.21 \\
     $\log_{10} A_{\rm SOBHB}$ & $-12.38$ & 0.34 \\
     $\log_{10} A_{\rm ExWD}$  & $-11.06$ & 0.17 \\
     $\log_{10} A_{\rm EMRI}$  & $-11.34$ & 0.5  \\
     \hline
    \end{tabular}
    \caption{Fiducial values of the different foreground log-normalizations. The EMRI amplitude is quoted at the LISA reference frequency $3\,{\rm mHz}$, with the tabulated spectrum of Ref.~\cite{Perego:2025bif} normalized accordingly. The MBHB contribution is taken into account in the analysis, but fixed to its
    tabulated data [see explanation below Eq.~(\ref{eq:mbhb_template})].}\label{tab:foreground_fiducial}
\end{table}

We note that in contrast to the other foreground components, we will treat the MBHB contribution as a fixed foreground (taken directly from the tabulated data), and no inference will be performed over this signal. This is because its shape is highly sensitive to the %exact 
population properties, while its contribution is expected to be marginal in our case of study. In Ref.~\cite{Perego:2025bif} two population hypotheses are considered (HS and LS), and the resulting foregrounds are quite different. However, both of them have sub-leading amplitudes versus the other foregrounds in the frequency range of interest for LISA, $f \gtrsim 10^{-4}$ Hz. Thus, even if, for concreteness, we use the HS hypothesis in our analysis,  considering either of them is expected to be only marginally affecting our results, as the only impact they might have is at low frequencies, $f < 10^{-4}$ Hz, where their power rise, but LISA's sensitivity declines.  

Combining all five contributions, the total astrophysical foreground contribution present at LISA window reads,
\begin{equation}
\begin{aligned}
h^2\Omega_{\rm FG}(f)
&=
h^2\Omega_{\rm WD}(f)
+
h^2\Omega_{\rm SOBHB}(f)
+
h^2\Omega_{\rm ExWD}(f)
\\
&\quad
+
h^2\Omega_{\rm EMRI}(f)
+
h^2\Omega_{\rm MBHB}(f).
\end{aligned}
\end{equation}

\begin{figure*}[!t]
\centering
\includegraphics[width=0.7\textwidth]{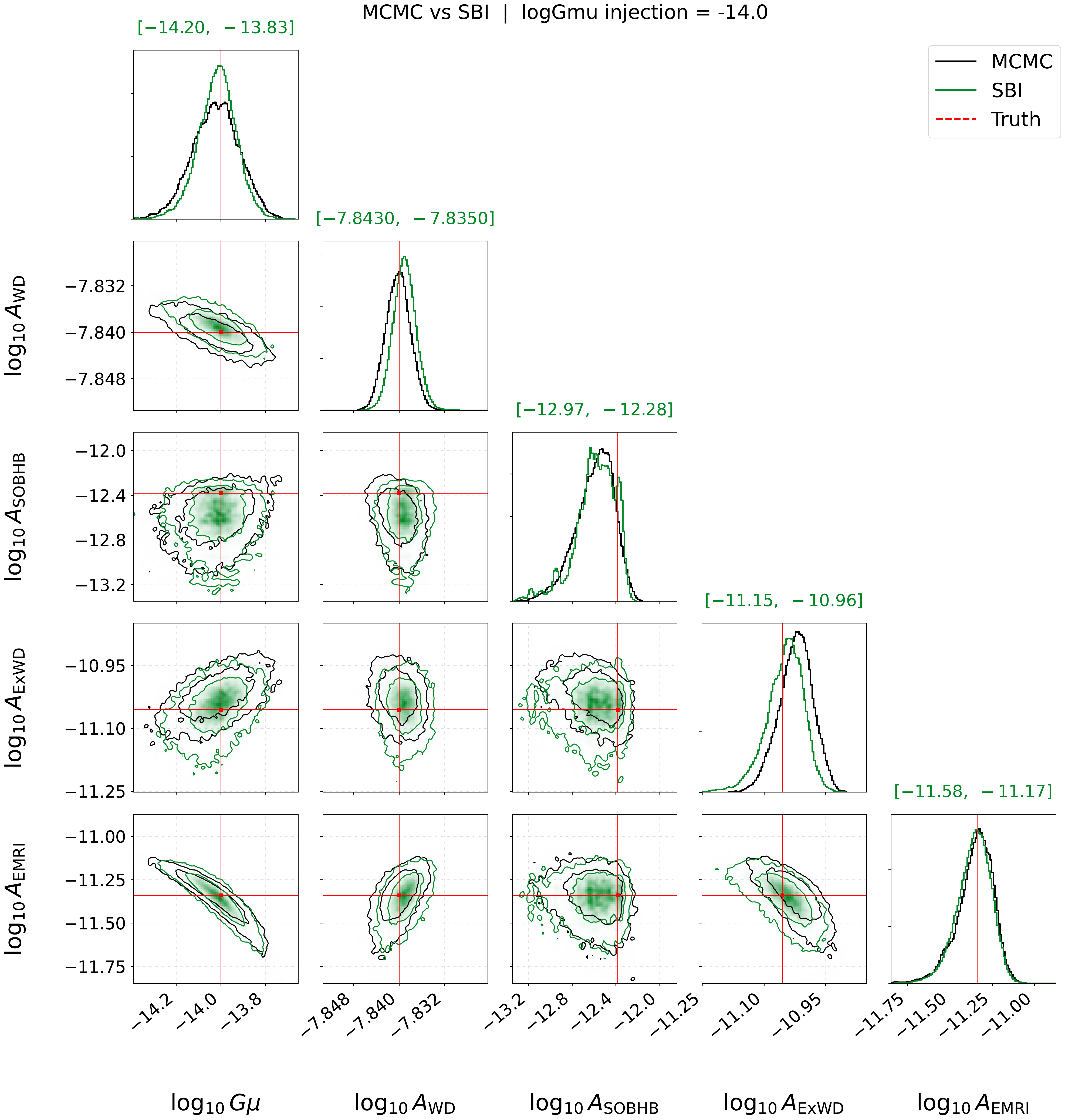}\\[-1em]
\caption{\label{fig:corner_plot}
An example of marginalized posterior distributions for a single mock LISA
observation with astrophysical foregrounds included, comparing the
SBI posterior (green) to an independent MCMC analysis (black)
at an injected string tension $G\mu_{\rm inj}=10^{-14}$. Red lines
indicate the injected parameter values. The interval on top of each column denotes the 95\% highest-density interval of each parameter's posterior.
The predictions on the
instrumental-noise parameters ($A_{\rm acc}$ and $A_p$) are omitted for
clarity, but they have been jointly inferred together with the rest of
the parameters.
}
\end{figure*}

Apart from MBHBs, we will infer all foreground parameters $\lbrace A_{\rm WD}, A_{\rm SOBHB}, A_{\rm ExWD}, A_{\rm EMRI} \rbrace$ simultaneously, together with the LISA instrumental noise parameters $A_{\rm acc}, A_p$, and the cosmic-string
tension $G\mu$.

We collect in Table \ref{tab:foreground_fiducial} the mean foreground amplitudes (each of them normalized to different pivot frequencies), from now on the `foreground fiducial model', as well as the prior ranges corresponding to the foreground parameters jointly entering the inference. 

\section{Results}
\label{sec:results}
As in {\tt Paper\,\,I}, our statistical approach is based on the Neural Posterior Estimation (NPE) flavor of Simulation-Based Inference (SBI), which we detail in Appendix~\ref{app:SBI}. The LISA noise modeling and data generation procedure are also discussed in Appendix~\ref{app:lisa_noise}. In this section we directly focus on the results of our analysis.

\begin{figure}[t]
\centering
\includegraphics[width=0.48\textwidth]{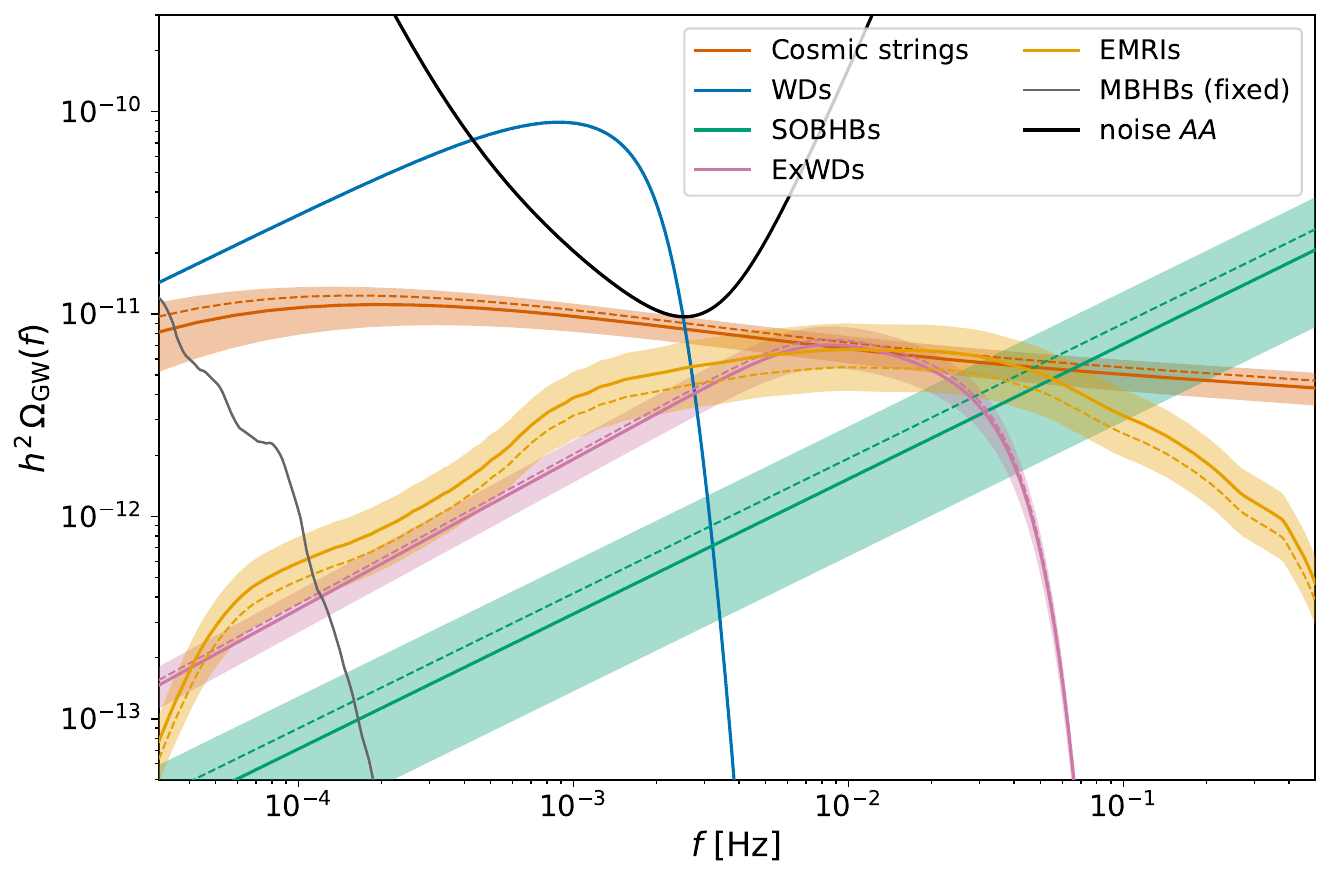}\\[-1.5em]
\caption{
Reconstructed spectra for the cosmic string signal of $G\mu_{\rm inj} = 10^{-14}$ (red) and the different foregrounds considered in this work (except the MBHB which is not reconstructed): Galactic binaries (WD, blue), stellar-origin Black-hole binaries (SOBHBs, green), extragalactic white dwarfs (ExWDs, pink), extreme-mass-ratio-
inspirals (EMRIs, yellow), and massive black-hole binaries (MBHB, gray), as well as LISA instrumental noise (AA channel). We show the injected contributions (dashed), posterior median (solid), as well as the 95\% CI bands.}
\label{fig:spectra_posterior}
\end{figure}

An example of
posterior is shown 
in the corner plot of
Fig.~\ref{fig:corner_plot}, where we compare our SBI results with independent MCMC posteriors for a representative injected string tension,
$G\mu_{\rm inj}=10^{-14}$. Two additional examples, for 
tensions
$G\mu_{\rm inj}=10^{-12}$ and $10^{-13}$, are shown in
Appendix~\ref{app:more_fid}. As the two methods, SBI and MCMC, give compatible results, we interpret this as a first validation of the amortized inference framework in the presence of astrophysical foregrounds. We discuss this in more detail in {\it Robustness of results} (see below), where we further validate our SBI output via coverage tests. We note that, by virtue of amortization, the same trained network can be applied to a large ensemble of independent mock observations to quantify reconstruction performance and posterior calibration;  something not possible with MCMC.

From the joint posterior distribution, we can generate predictions for the spectra of cosmic strings, foregrounds, and LISA noises. The spectrum reconstruction for $G\mu_{\rm inj}=10^{-14}$ is shown in Fig.~\ref{fig:spectra_posterior}, for the fiducial foregrounds and noise modelings.  
For all spectra for which we make inference, ({\it i.e.}~all except for the MBHB foreground), 
we show the true injected value (dashed), the posterior median (solid), and the  95\% credible interval band. The reconstruction of a cosmic-string signal with $G\mu_{\rm inj}=10^{-14}$ 
is more precise than the reconstructions of the SOBHB, ExWD and EMRI foregrounds. The WD foreground is however reconstructed more precisely, thanks to its large amplitude above other contributions. 

\begin{figure*}[!tp!]
\centering
\includegraphics[width=0.85\textwidth]
{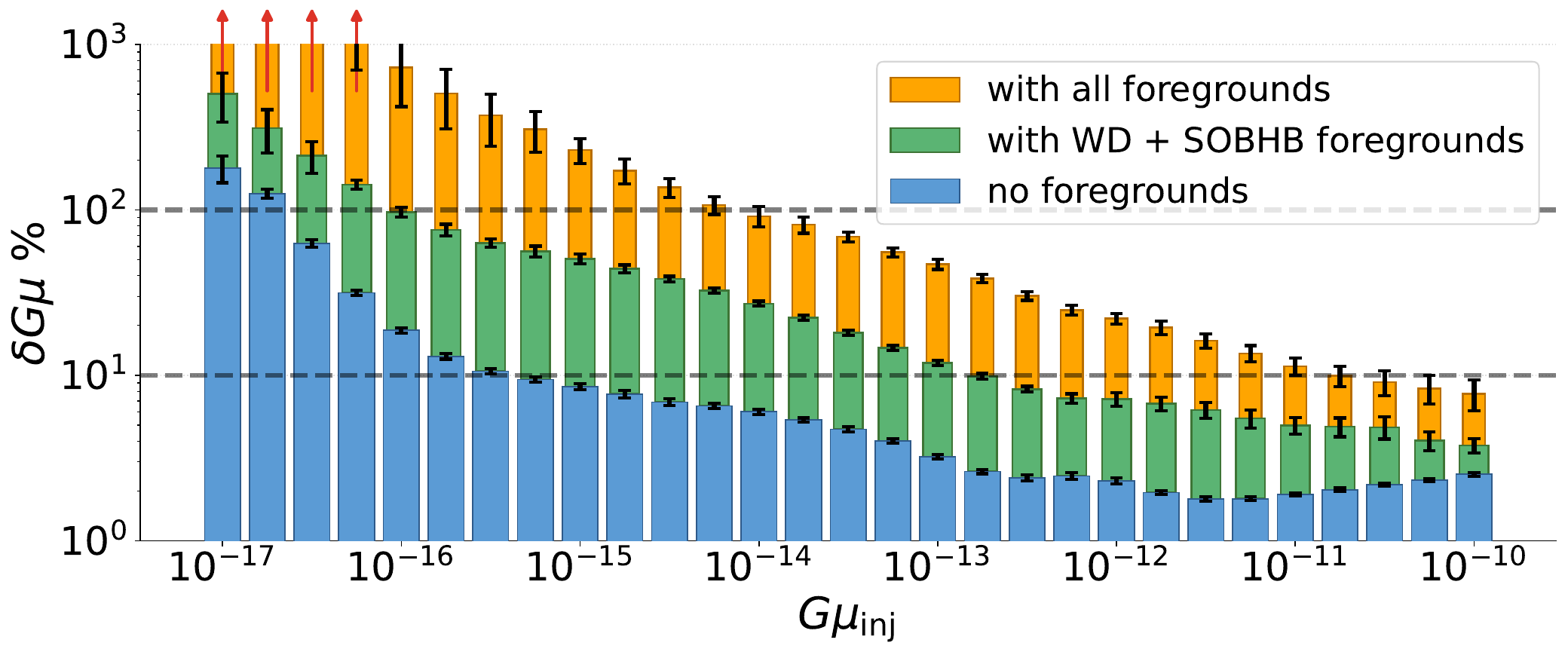}
\includegraphics[width=0.85\textwidth]{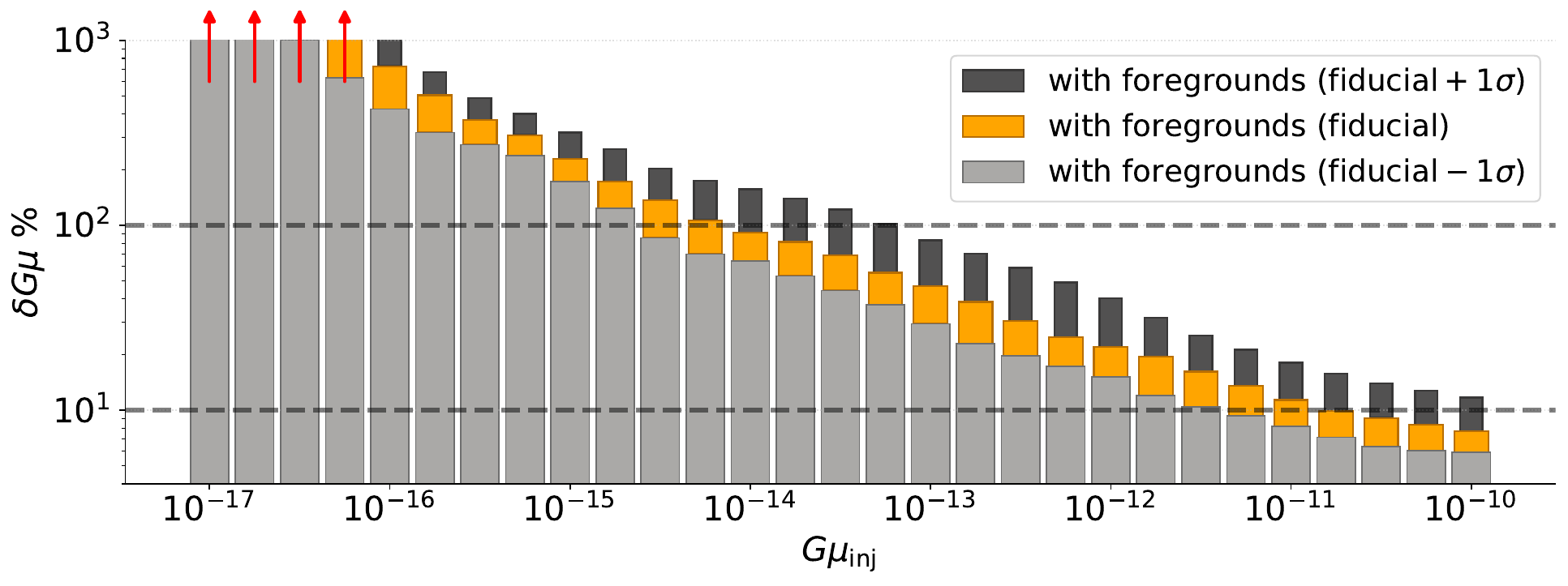}
\caption{Reconstruction precision $\delta G\mu$\,$[\%]$ as defined in Eq.~\eqref{eq:precision}, as a function of the injected string tension $G\mu_{\rm inj}$. In the \emph{Upper Panel} we assume three different foreground budgets: blue bars show the foreground-free precision; green bars indicate the degradation due to simultaneously inferring the astrophysical foreground amplitudes when assuming only two contributions (WDs and SOBHBs); and orange bars show the degradation when we assume instead our fiducial foreground model containing all five contributions (SOBHBs, WDs, ExWDs, EMRIs, MBHBs), see Table \ref{tab:foreground_fiducial}. The error bars indicate the standard deviation among 20 data realizations. In the \emph{Lower Panel} we consider three foreground-amplitude scenarios, all of which include all five foreground contributions: the fiducial values (orange), a conservative case where these fiducials are shifted by $+1\sigma$ (dark gray) according to their corresponding priors, and an ``optimistic'' case where the fiducials are shifted by $-1\sigma$ (light gray).  The red arrows in both panels indicate that the reconstruction precision exceeds the plot range.}
\label{fig:precision_normal}
\end{figure*}

\begin{figure}[t]
\centering
\includegraphics[width=0.45\textwidth
,height=7cm]
{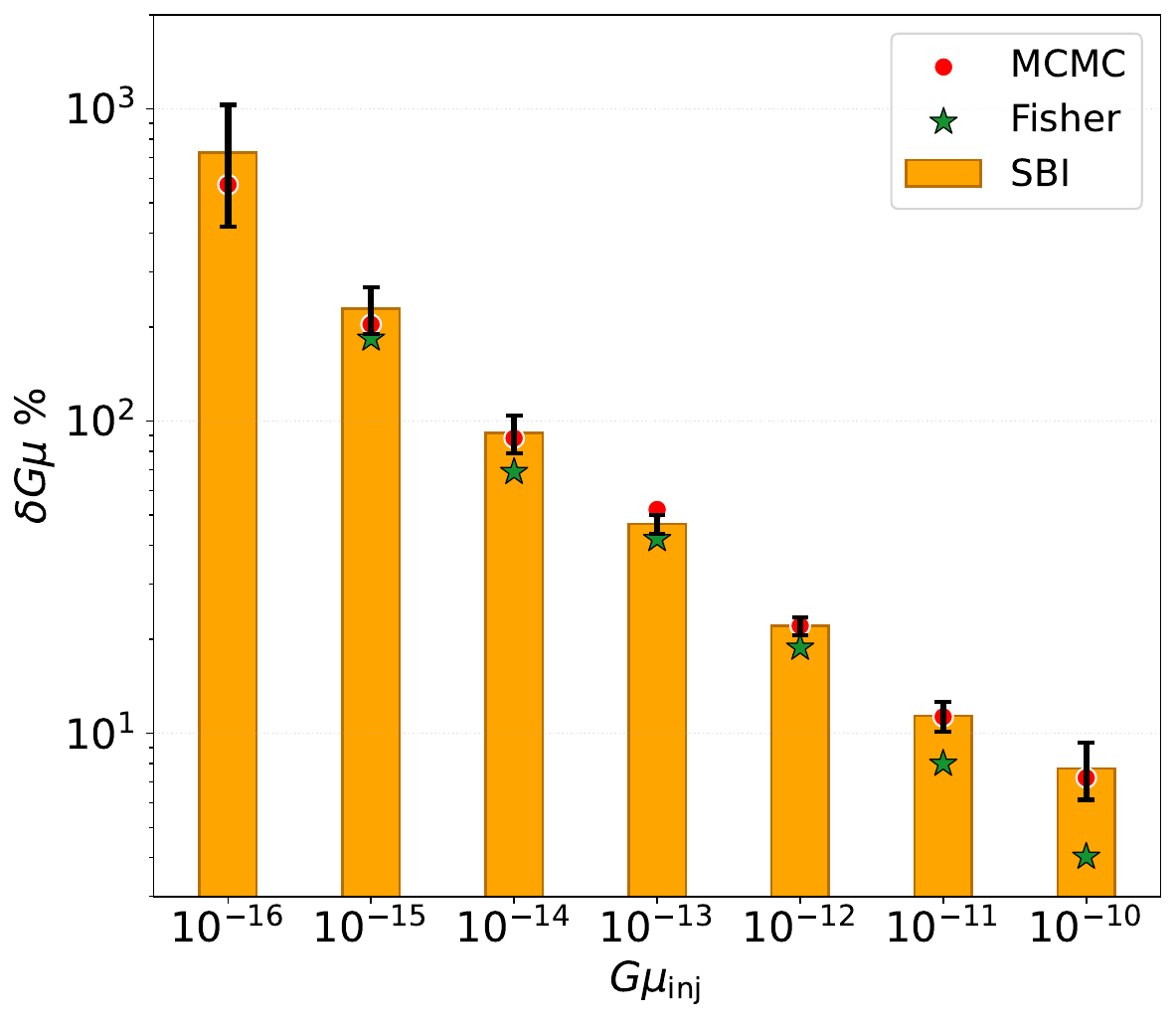}\\[-1.5em]
\caption{Foreground-aware reconstruction precision from SBI (orange bars) compared with independent MCMC results (red circles) and Fisher-matrix forecasts (green stars). Error bars denote the standard deviation in SBI analysis across 20 independent mock LISA realizations. See also Table~\ref{tab:results}.
}
\label{fig:precision_comparison}
\end{figure}

%\vspace{0.5cm}
\textit{Reconstruction precision.---}To quantify the reconstruction of the cosmic-string tension, for each injected value $G\mu_{\rm inj}$ we generate $N_{\rm datasets}=20$ independent mock LISA realizations, perform SBI inference on each, and compute the $95\%$ highest-density interval (HDI) $[G\mu_{\rm lo},G\mu_{\rm hi}]$, of the posterior on $G\mu$. Following {\tt Paper\,\,I}~\cite{Dimitriou:2025bvq}, we define the 95\%-{\it reconstruction precision} as
\begin{equation}
\label{eq:precision}
\delta G\mu\, [\%] \equiv
\left(\frac{G\mu_{\rm hi}-G\mu_{\rm lo}}
{G\mu_{\rm inj}}\right) \times 100 \,.
\end{equation}
We plot this quantity as a function of the injected string tension $G\mu_{\rm inj}$ in Fig.~\ref{fig:precision_normal} and Fig.~\ref{fig:precision_comparison}, for the fiducial noise modeling ({\it c.f.}~Appendix \ref{app:lisa_noise}). For reference, we also list our SBI reconstruction results in \% in Table~\ref{tab:results}, where we also compare them to analogous reconstructions via independent MCMC and {\it Fisher} analysis. Both Figs.~\ref{fig:precision_normal} and~\ref{fig:precision_comparison}, together with  Table~\ref{tab:results}, represent the main results of this paper, which we discuss now in detail. 
 
\textit{Effects of foregrounds.}---The impact of astrophysical foregrounds on the reconstruction of the cosmic-string tension is shown in the top panel of Figure~\ref{fig:precision_normal}. Without astrophysical foregrounds, the reconstruction precision (blue bars) reaches the few-percent level for $G\mu_{\rm inj}\gtrsim10^{-12}$, in particular $\sim 2.3\%$ at $10^{-12}$ and $\sim 2.5\%$ at $10^{-10}$, similarly as the results presented previously in {\tt Paper\,\,I}. Once all five known astrophysical foregrounds (SOBHBs, WDs, ExWDs, EMRIs, MBHBs) are included, the reconstruction (orange bars) is substantially degraded across the entire parameter range. If we consider only as foreground contributions SOBHBs and WDs, as {\it e.g.}~in~\cite{Blanco-Pillado:2024aca}, the reconstruction (green bars) is also degraded, though not as much. We postpone a discussion on this reduced foreground budget case for later, and focus now on the comparison between foreground-free and full-budget foreground cases. 

The inclusion of all foregrounds reduces the range of string tensions that can be reconstructed with useful precision. Taking $\delta G\mu<100\%$ as a threshold for meaningful reconstruction, the minimum {\it reconstructible} tension shifts from approximately $G\mu \simeq 2.1\cdot 10^{-17}$ in the absence of foregrounds, to $G\mu \simeq 7.2\cdot10^{-15}$ in the presence of foregrounds, corresponding to a degradation of roughly $2.5$ orders of magnitude in
string tension. For a very good reconstruction, capable of confidently distinguishing the signal from the noise, we adopt a threshold of $\delta G\mu<10\%$. Under this stricter criterion, the minimum 
tension required shifts from $G\mu\simeq 4.2\times10^{-16}$ in the absence of foregrounds, to $G\mu\simeq 1.7\times10^{-11}$ when foregrounds are included, corresponding to a degradation of roughly $4.6$ orders of magnitude in 
string tension---nearly double the shift seen under the $100\%$ threshold criterion. We conclude therefore that foregrounds impact high-confidence detections even more severely than marginal detections. 

Another way to quantify this, is to define a {\it foreground excess} factor as the ratio of the foreground-aware to the foreground-free precision. This factor decreases smoothly from approximately $34$ at $G\mu_{\rm inj}=10^{-17}$ to $27$ at $10^{-15}$, $9.6$ at $10^{-12}$, and $3.1$ at $10^{-10}$. The relative impact of the foregrounds is therefore strongest for weak cosmic-string signals, where the cosmological background lies well below the astrophysical foreground, and gradually decreases as the signal becomes increasingly dominant.

\begin{table}[t!]
\centering
\begin{tabular}{|c|ccc|c|c|}
\cline{1-4} \cline{6-6}
\multirow{2}{*}{~$G\mu_{\rm inj}$~} & \multicolumn{3}{c|}{~$\delta G\mu$ [\%] with foregrounds~}               &  & \multicolumn{1}{c|}{\begin{tabular}[c]{@{}c@{}}\\[-0.3cm] ~ $\delta G\mu$ [\%] (SBI) ~\end{tabular}}\\ \cline{2-4}  
                     & \multicolumn{1}{c|}{\begin{tabular}[c]{@{}c@{}}\\[-0.3cm] ~ ~SBI~ ~\end{tabular}}   & \multicolumn{1}{c|}{\begin{tabular}[c]{@{}c@{}}\\[-0.3cm] ~ MCMC ~\end{tabular}} & \multicolumn{1}{c|}{\begin{tabular}[c]{@{}c@{}}\\[-0.3cm] ~ Fisher ~\end{tabular}} &  & no foreground   \\ \cline{1-4} \cline{6-6} 
\multicolumn{1}{|c|}{\begin{tabular}[c]{@{}c@{}}\\[-0.3cm] $10^{-16}$\end{tabular}}                   & \multicolumn{1}{c|}{\begin{tabular}[c]{@{}c@{}}\\[-0.3cm] 725.47\end{tabular}}   & \multicolumn{1}{c|}{\begin{tabular}[c]{@{}c@{}}\\[-0.3cm] 570.12\end{tabular}} & \multicolumn{1}{c|}{\begin{tabular}[c]{@{}c@{}}\\[-0.3cm] $1.85 \times 10^5$\end{tabular}} &  & \multicolumn{1}{c|}{\begin{tabular}[c]{@{}c@{}}\\[-0.3cm] 18.71\end{tabular}}               \\
$10^{-15}$                    & \multicolumn{1}{c|}{229.49}   & \multicolumn{1}{c|}{200.12}    & 183.78      &  & 8.56               \\
$10^{-14}$                    & \multicolumn{1}{c|}{91.47}   & \multicolumn{1}{c|}{88.23}    & 68.73      &  & 6.03               \\
$10^{-13}$                    & \multicolumn{1}{c|}{46.88}   & \multicolumn{1}{c|}{52.11}    & 41.90      &  & 3.22               \\
$10^{-12}$                    & \multicolumn{1}{c|}{22.06}   & \multicolumn{1}{c|}{22.10}    & 18.81      &  & 2.30               \\
$10^{-11}$                    & \multicolumn{1}{c|}{11.36}   & \multicolumn{1}{c|}{11.30}    & 7.99      &  & 1.91               \\
$10^{-10}$                   & \multicolumn{1}{c|}{7.77}   & \multicolumn{1}{c|}{7.26}    & 4.02      &  & 2.52               \\ \cline{1-4} \cline{6-6} 
\end{tabular}
\caption{
Reconstruction precision in \%, defined in Eq.~\eqref{eq:precision}, for different injected values of the cosmic-string tension. The first set of results on the left table include foregrounds in the analysis, where we compare our SBI results to MCMC's and Fisher's. The right table is the reconstruction precision when foregrounds are omitted.
}
\label{tab:results}
\end{table}

To gauge the dependence of the above conclusions on the assumed foreground modeling, we repeat an analogous analysis, but shifting all foreground amplitudes simultaneously by $\pm1\sigma$ relative to the fiducial choices ({\it i.e.}~the mean values of the foreground amplitude priors). The resulting precisions are shown in the bottom panel of Fig.~\ref{fig:precision_normal}. As expected, decreasing the foreground amplitudes by $-1\sigma$, systematically improves the reconstruction, while increasing them by $+1\sigma$, further degrades it. At $G\mu_{\rm inj}=10^{-17}$ the reconstruction precision ranges from $3.1\times10^3\, \%$ in the optimistic (low-foreground) scenario, to $6.1\times10^3 \, \%$ for the fiducial model and almost $1.0\times10^4 \,\%$ for the pessimistic (high-foreground) scenario, {\it i.e.}~exhibiting very poor reconstruction in all cases. At $G\mu_{\rm inj}=10^{-12}$ the corresponding precisions are, respectively, $15.2\%$, $22.1\%$, and $40.3\%$, while at $10^{-10}$ they reduce to $5.9\%$, $7.7\%$, and $11.7\%$. The reconstruction threshold $\delta G\mu \leq 100\%$, similarly requires variation from approximately $G\mu\simeq2.5\times10^{-15}$ in the optimistic scenario, to $G\mu\simeq5.9\times10^{-14}$ in the pessimistic scenario. Requiring instead that $G\mu$ can be reconstructed precisely with $\delta G\mu\lesssim 10\%$ uncertainty, pushes the region that can be probed from $G\mu \gtrsim 1.74\times10^{-11}$ for the fiducial model, to $G\mu \gtrsim 3.9\times10^{-12}$ for the optimistic scenario, and $G\mu \gtrsim 2.3\times10^{-10}$ for the pessimistic scenario. 

To summarize, in the presence of all expected foregrounds, for the signal to be well reconstructible, an increment of $4$--$6$ orders of magnitude in the string tension is required. Even with the most optimistic foreground modeling (at $-1 \sigma$ away from prior means), the ability of LISA to reconstruct the cosmic-string tension remains substantially degraded as compared to the foreground-free case. This suggests that the dominant limitation arises from the presence of --- and potential degeneracies with --- all relevant astrophysical foregrounds in the LISA band, as we discuss next.

\textit{Comparison to a reduced foreground budget.---} We note that analogous reconstructions of the tension of cosmic-string models, considering only the contributions from SOBHBs and WDs in the foreground budget, were presented in Refs.~\cite{Boileau:2021gbr} and~\cite{Blanco-Pillado:2024aca}, with the latter including the BOS modeling. Our previous results, which on top of the SOBHB and WD foregrounds, include also the contributions from MBHBs, ExWDs and EMRIs, differ in fact significantly. A direct quantitative comparison with Refs.~\cite{Boileau:2021gbr,Blanco-Pillado:2024aca} cannot be done straightforwardly, however, due differences in the analysis. To compare with a situation where only SOBHB and WD contributions are considered in the foreground budget, we have therefore obtained our own reconstruction of the tension assuming such setup (see green bars in the top panel of Fig.~\ref{fig:precision_normal}), and compared it against our previous full-budget foreground analysis (orange bars in the same panel).  

We find that a reconstruction threshold of $\delta G\mu \lesssim 10\%$ requires $G\mu \gtrsim 1.7\cdot 10^{-13}$ when only SOBHB and WD foregrounds 
are considered, versus the $G\mu \gtrsim 1.7\cdot 10^{-11}$ requisite from the full-foreground budget analysis. Loosening the reconstruction threshold to $\delta G\mu \lesssim 100\%$, requires tensions as $G\mu \gtrsim 1.0\cdot 10^{-16}$ in the SOBHB+WD foreground scenario, which are much smaller than the values $G\mu \gtrsim 7.2\cdot 10^{-15}$ needed in the full-foreground budget scenario. Essentially, to meet the same reconstruction threshold, the required signal corresponds to a string tension $\sim$1--2 orders of magnitude larger in the full-foreground budget case, than in the SOBHB+WD foreground only scenario.

Our results indicate, therefore, that it is necessary to include all known foregrounds within the LISA band, in order to assess the reconstruction error of cosmic-string signals. In particular, the ExWD and EMRI foregrounds are responsible for the higher degradation level of the tension reconstruction, as these two foregrounds cover a significant part of the LISA band in the frequencies just right-next-to the galactic WD foreground, with amplitudes way above the SOBHB foreground, see Fig.~\ref{fig:spectra_posterior}.

\textit{Robustness of results.}---In order to verify the robustness of our SBI results in the full-budget foreground case, we compare some cases against MCMC and Fisher techniques. As Fisher is simply an approximation to data analysis, our SBI results are not necessarily expected to agree well with it. To validate our SBI results, instead, a better agreement with MCMC is expected, given that for the latter we adopted a likelihood which has proven to be sufficiently reliable (see Appendix \ref{app:mcmc_likelihood}). 

From Table \ref{tab:results} we see that MCMC agrees with SBI results at a ${\rm few}~\%$ level for all tension values, except for the smallest one, $G\mu=10^{-16}$, where precisions disagree at $\sim 20\%$. This discrepancy is however still consistent with the statistical fluctuations between realizations, as it lies well within the realization-to-realization scatter indicated by the error bars in Fig.~\ref{fig:precision_comparison}.

On the other hand, the agreement between Fisher and SBI is not as good, as expected, shifting roughly from around $10\%$ to $50\%$ for tensions above $G\mu \gtrsim 10^{-15}$. For small tensions the Fisher precision deteriorates completely, as shown {\it e.g.} for $G\mu=10^{-16}$ in Table~\ref{tab:results}. For sufficiently small tensions, the signal derivative becomes degenerate with nuisance directions, so changes in $G\mu$ can be locally--- at the likelihood level --- absorbed by foreground/noise variations. In our case this occurs for $G\mu \ll 10^{-15}$, with the Fisher precision becoming larger than the SBI prediction for $\log_{10} G\mu\lesssim-15.5$.

Internal posterior calibration of our method has been also assessed using coverage checks. For both the foreground-free and full-budget foreground models, we draw parameter vectors from the corresponding priors, simulate the associated mock LISA observations, and compute the empirical coverage as a function of the nominal credibility level for every inferred parameter. We have verified that out posteriors are well calibrated, see Appendix~\ref{app:coverage} for details and coverage plots. These results confirm that the inclusion of astrophysical foreground nuisance parameters does not compromise the statistical calibration of the amortized SBI framework. We also highlight that the amortized nature of the SBI approach adopted in this work, allows to run coverage analyses involving thousand of inferences on simulated data sets.

\section{Conclusions}
\label{Conclusions}

Cosmic defect networks, and in particular cosmic strings, can be probed through their imprints on the cosmic microwave background, via 2- and 3-point correlation functions~\cite{Garcia-Bellido:2010qjz,Figueroa:2010zx,Fenu:2013tea,Ade:2013xla,Durrer:2014raa,Lizarraga:2014xza,Charnock:2016nzm,Lizarraga:2016onn,Lopez-Eiguren:2017dmc,Ringeval:2010ca,Regan:2014vha}. In this work, we instead quantify the ability of planned space-based interferometric detectors such as LISA, to reconstruct the GWB signal from cosmic-string networks. Using simulation-based inference (SBI) techniques, we quantify the reconstruction precision of conventional cosmic-string GWBs at LISA in the presence of known leading astrophysical foregrounds expected in the LISA band. We focus on the standard Nambu--Goto limit of local strings propagating in a $\Lambda$CDM cosmology, described by the BOS template in Ref.~\cite{Dimitriou:2025bvq}. This choice corresponds to an optimistic scenario, as the signal in this modeling corresponds to one of the largest GWBs that can be expected naturally in an early Universe scenario.

New to the present analysis, compared to previous works~\cite{Boileau:2021gbr,Blanco-Pillado:2024aca}, is the inclusion of foregrounds of extragalactic white-dwarf binaries (ExWD), extreme mass-ratio inspirals (EMRIs), and massive black hole binaries (MBHBs), in addition to the previously considered foregrounds of stellar-origin black-hole binaries (SOBHBs), and Galactic white-dwarf binaries (WDs). We have performed a joint inference over the cosmic-string tension $G\mu$, the LISA instrumental-noise parameters $A_{\rm acc}$ and $A_p$ (see Appendix~\ref{app:lisa_noise} for definition), and the astrophysical foregrounds amplitudes $A_{\rm WD}$, $A_{\rm SOBHB}$, $A_{\rm ExWD}$, and $A_{\rm EMRI}$,   see Table~\ref{tab:foreground_fiducial} (the MBHB contribution was included as a fixed component, but its impact on the signal reconstruction at LISA is rather marginal, given its spectral shape). Using this setup, we assessed how the detectability of the cosmic-string GWB is affected by the presence of these dominant astrophysical foregrounds. 

Our main result is that inclusion of astrophysical foregrounds substantially degrades the reconstruction of the cosmic-string tension $G\mu$. In comparison to parameter reconstruction without foregrounds, the inclusion of foregrounds degrades the precision reconstruction of $G\mu$ by over an order of magnitude for most tensions. 
To reconstruct $G\mu$ at a precision $\delta G\mu<100\%$, the minimum reconstructible tension shifts from approximately
$G\mu\simeq{2.1\times10^{-17}}$ in the foreground-free case to $G\mu\simeq{2.5\times10^{-15}}$ ($G\mu\simeq{7.2\times10^{-15}}$) [$G\mu\simeq5.9\times10^{-14}$] for the optimistic (fiducial) [pessimistic] foreground model. Requiring instead a reconstruction at the level of
$\delta G\mu\lesssim 10\%$, pushes the accessible region from $G\mu \gtrsim{4.2\times10^{-16}}$ to
$G\mu \gtrsim{3.9\times10^{-12}}$ ($G\mu \gtrsim1.74\times10^{-11}$) [$G\mu \gtrsim2.3\times10^{-10}$] correspondingly.
%for the fiducial (optimistic) [pessimistic] foreground model. 
In the presence of the expected astrophysical foregrounds at LISA, our results imply a dramatic $4-6$ orders of magnitude increase in tension strength for a signal to be {\it reconstructible}. 

As a consistency check,
we compared our 
SBI posteriors with 
MCMC and Fisher analyses. The SBI and MCMC results agree well over the range of injected tensions, 
supporting the use of SBI. Fisher forecasts give comparable (but systematically smaller) results at sufficiently large tensions, and become unreliable in the low-signal regime. We also verified posterior calibration using coverage tests, finding empirical coverages consistent with the nominal credible levels for both the foreground-free and full-budget foreground-aware analyses. 

Our results show that LISA remains sensitive to conventional cosmic-string backgrounds at sufficiently large values of the string tension $G\mu$, but 
the reconstruction precision of the signal, and hence of the tension, is substantially hampered when a comprehensive state-of-the-art astrophysical foreground budget is included.
Regarding other cosmic-string models (see classification in {\tt Paper\,\,I}), we expect that their precise reconstructions, already substantially degraded by degeneracies among model parameters~\cite{Dimitriou:2025bvq}, would be even more challenging due to foregrounds. Accurate foreground understanding becomes therefore essential for robust quantification of the ability of LISA to measure cosmic strings, and more generally to measure any other cosmological background.

\begin{acknowledgments}
We thank A.~Sesana and F.~Pozzoli for insightful discussion and assistance to estimate the uncertainty of the EMRI foreground, providing smooth versions of the physical strain. This work is supported by the grants PROMETEO/2021/083, CIDEGENT/2020/055, EUR2022-134028, PID2023-148162NB-C22, PRTR-C17.I01, ASFAE/2022/020, ASFAE/2022/024, and CIPROM/2022/69. 
\end{acknowledgments}

\appendix

\section{Additional reconstruction cases}
\label{app:more_fid}

\begin{figure*}[t]
\centering
\includegraphics[width=0.48\textwidth]{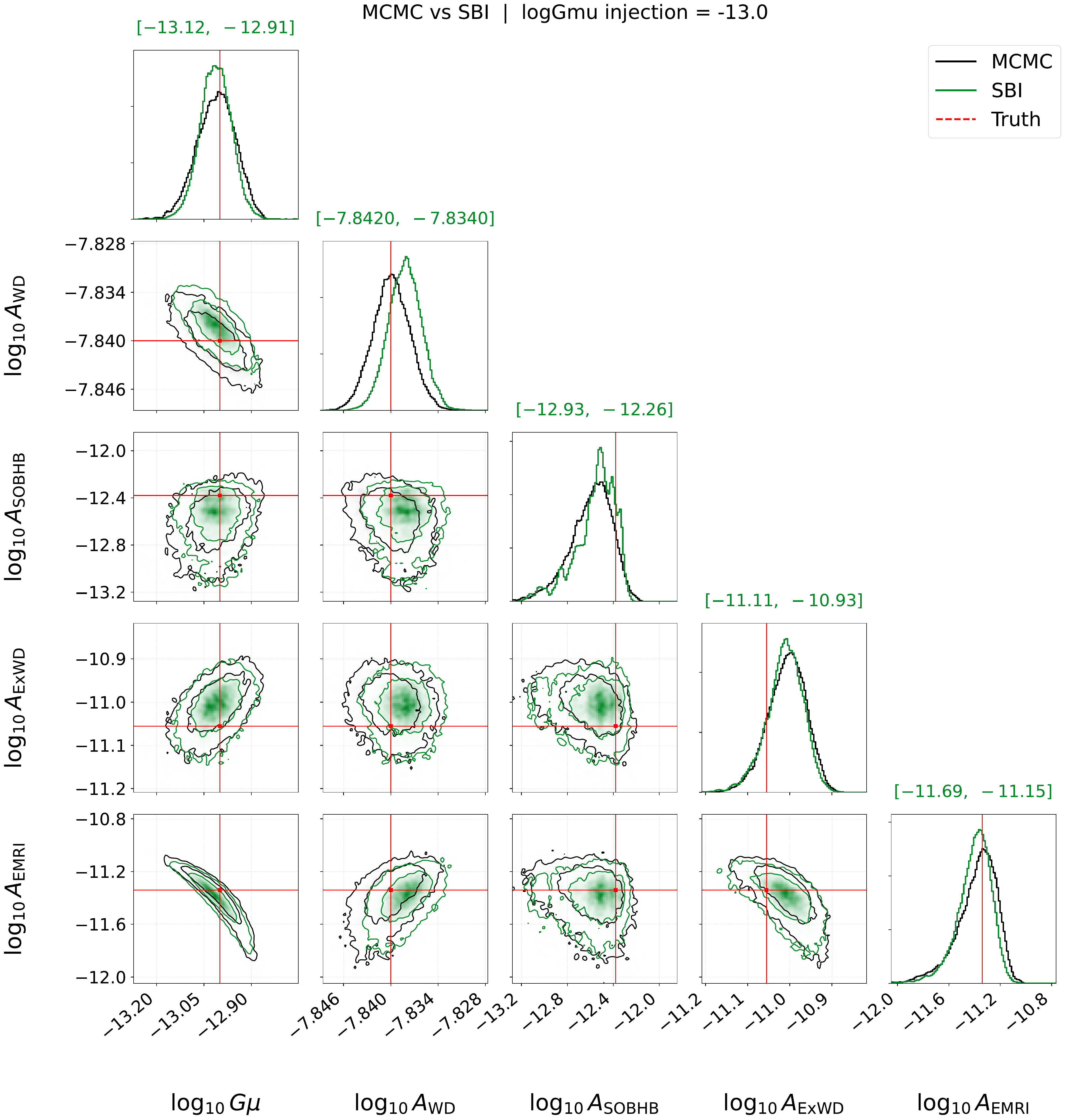}
\hfill
\includegraphics[width=0.48\textwidth]{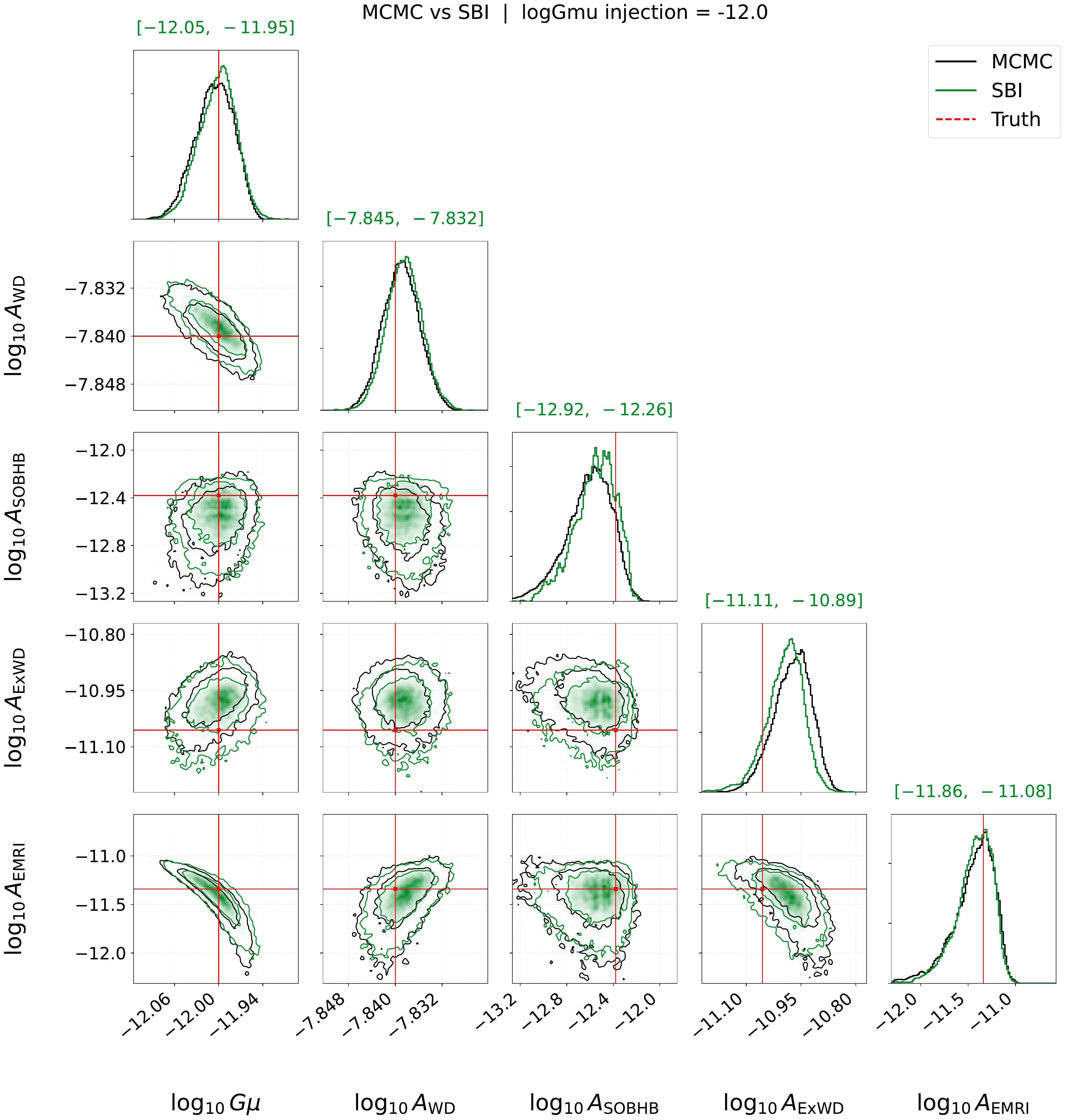}
\caption{%\label{fig:corner_plot}
Example marginalized posterior distributions for two single mock LISA observations with astrophysical foregrounds included, comparing the SBI posterior (green) to an independent MCMC analysis (black) at injected string tensions $G\mu_{\rm inj}=10^{-13}$ (left) and $10^{-12}$ (right). Red lines indicate the injected parameter values. The interval on top of each column denotes the 95\% highest-density interval of each parameter's posterior. The prediction on the instrumental-noise parameters ($A_{\rm acc}$ and $A_p$) are omitted for clarity, but they have been jointly inferred together with the rest of parameters. 
}
\label{fig:more_mcmc_fid}
\end{figure*}

Here we complement the results presented in
Sec.~\ref{sec:results} with two additional examples of single-realization
posteriors, for injected string tensions $G\mu_{\rm inj}=10^{-13}$ and
$10^{-12}$. As in the main text, each mock LISA observation includes the
instrumental noise and the full set of astrophysical foregrounds, and we
compare the posterior obtained from SBI against an
independent MCMC analysis of the same data realization. The resulting
marginalized posteriors are shown in Fig.~\ref{fig:more_mcmc_fid}. In both
cases the two methods yield compatible posteriors, with consistent
degeneracy structures among the foreground amplitudes, further supporting
the validity of the amortized inference framework across the range of
injected tensions. 

\section{LISA noise model and data generation}
\label{app:lisa_noise}
For completeness, we summarize here the LISA noise model and mock-data generation procedure used throughout this work; we refer the reader to App.~D of {\tt Paper\,\,I}~\cite{Dimitriou:2025bvq} for the full derivation.

After time-delay interferometry, 
the instrumental noise budget reduces to 2 effective contributions: the interferometry metrology system (IMS) and acceleration noises, with power spectral densities $P_{\rm IMS}(f;A_P)$ and $P_{\rm acc}(f;A_{\rm acc})$, parametrized by the amplitudes $A_P$ and $A_{\rm acc}$. We work in the uncorrelated $A$, $E$, $T$ TDI basis, for which the noise power spectra $N_{AA}=N_{EE}$ and $N_{TT}$ are obtained from $P_{\rm IMS}$ and $P_{\rm acc}$, 
and converted to an equivalent energy-density spectrum $\Omega_{\rm noise}^{\alpha\beta}(f;A_P,A_{\rm acc})$ via the LISA response functions $\tilde R_{\alpha\beta}(x_f)$, with $x_f\equiv 2\pi f L/c$ and $L=2.5\times10^9\,$m the LISA arm length. We assume uniform priors on $A_P$ and $A_{\rm acc}$, centered at their fiducial values $A_P=15$ and $A_{\rm acc}=3$ with a $20\%$ margin.

Mock data are generated over a $3\,{\rm yr}$ science run, segmented into $N_c=94$ chunks of $11.5$ days each. Within each chunk and frequency bin, the data $D^{\alpha\beta}_{i,j}=S_{i,j}+\mathcal{N}^{\alpha\beta}_{i,j}$ are built as the sum of independent Gaussian-distributed signal and noise realizations, following the standard cross-correlated-estimator construction for a stochastic background~\cite{Dimitriou:2025bvq}, and then averaged over the $N_c$ chunks to give $\bar D^{\alpha\beta}_i$. To reduce computational cost we coarse-grain the data above $\delta f/f\ll1$, re-binning the range $[10^{-3},0.5]\,$Hz into $1000$ logarithmically spaced macro-bins while retaining the original fine bins below $10^{-3}\,$Hz, for a total of $1970$ bins per chunk; the coarse-grained frequencies and data are obtained as noise-weighted averages within each macro-bin, with weights $w_i\propto[\Omega_{\rm noise}(f_i;\mathbf{n})]^{-1}$. The quantity $n_k$ appearing in Eqs.~\eqref{eq:loglike_gauss}--\eqref{eq:loglike_lognormal} is the number of fine-resolution bins contained within the $k$-th coarse macro-bin.

\section{Statistical approach}
\label{app:SBI}
We perform parameter inference using Neural Posterior Estimation (NPE) in the framework of simulation-based inference (SBI), which directly approximates posterior $p(\theta|x_\star)$ of the parameters $\theta$ given  data $x_\star$ from a set of simulated $(\theta,x)$ pairs, bypassing the need for a likelihood evaluation. In our implementation, the approximated posterior is parametrized as a neural network-based model (Normalizing Flow) whose training is 'armotized', i.e. not conditioned on the test data $x_\star$. We use the \texttt{sbi} package~\cite{tejero-cantero2020sbi}, in particular the implementation \texttt{NPE-C}. The parameter vector $\theta$ consists of the cosmic-string tension $\log_{10}G\mu$, the two LISA instrumental-noise parameters $A_{\rm acc}$ and $A_p$, and, when foregrounds are included, the four foreground amplitudes $\log_{10}A_{\rm WD}$, $\log_{10}A_{\rm SOBHB}$, $\log_{10}A_{\rm ExWD}$, and $\log_{10}A_{\rm EMRI}$ (note that the MBHB is taken into account, but fixed, so it is not part of the inference on $\theta)$. Priors are uniform on $\log_{10}G\mu \in [-18,-9]$, as well as on the noise parameters and on $A_{\rm acc}$, $A_p$ within $\pm 20\%$ of their fiducial values. The priors on the foreground log-amplitudes are quoted in Sect. \ref{sec:foregrounds}. 

For each draw of $\theta$ from the prior we generate a mock LISA data realization following the procedure of Sect.~\ref{sec:Backgrounds} and App.~\ref{app:lisa_noise}: the total signal $h^2\Omega_{\rm GW}(f)$ --- cosmic strings plus, where applicable, the astrophysical foregrounds --- is added to the instrumental noise PSDs in the $A$, $E$, $T$ TDI channels and averaged over $N_{\rm chunks}=94$ independent time segments. To reduce the data-load, LISA frequency bins within the range $\left[10^{-3}, 0.5\right] \mathrm{Hz}$ are re-binned onto 1000 logarithmically spaced frequency bins, while the small-frequency window $
\left[3\times 10^{-5},10^{-3}\right]$ Hz is not re-binned, making 970 bins with width $\Delta f=10^{-6}$ Hz. In total, thus, we work with $N_{\rm bins}=1970$ bins in the whole frequency window. The summary statistic fed to the network is the base-10 logarithm of the coarse-grained spectra in each channel:
\begin{equation}
{\bf x} = \left\{ \log_{10}D_{AA}(f_k),\; \log_{10}D_{EE}(f_k),\; \log_{10}D_{TT}(f_k) \right\},
\end{equation}
where $k=1,..,N_{\rm bins}$.
We generated two training sets: $2.5\times10^{5}$ simulations without foregrounds (three parameters) and $5\times10^{5}$ simulations including all foreground components (seven parameters). We use an independent NPE posterior model for each of these two setups. Training two separate amortized models in this way lets us isolate the degradation in $G\mu$ reconstruction caused specifically by the presence of astrophysical foregrounds, while keeping the LISA instrumental model identical in both cases. Once trained, a single forward pass of the network yields the full posterior for any observation; drawing $10^4$ posterior samples takes only a few seconds on a single node ($\sim 50$ CPU cores).
More details about the SBI implementation can be found in our previous work \cite{Dimitriou:2025bvq}.

\textit{Comparison against MCMC and Fisher forecasts.---}As a comparison of the amortized posteriors estimated from our SBI method, we additionally ran conventional MCMC sampling at a subset of injected values of $G\mu$, together with Fisher-matrix forecasts evaluated at the same fiducial points, using the mixed Gaussian+log-normal likelihood from our companion paper~\cite{Dimitriou:2025bvq} (see also Ref.~\cite{Flauger:2020qyi}); explicit expressions and implementation details are given in App.~\ref{app:mcmc_likelihood}. The comparison serves two purposes: the MCMC run provides an independent, well-calibrated cross-check of the SBI estimation---though, unlike our amortized NPE, it requires a fresh set of simulations for every new test dataset $x_\star$, making parameter scans far more costly---while the Fisher comparison quantifies the gap between the commonly adopted Fisher forecast and the more accurate MCMC/SBI posteriors. Since Fisher forecasts approximate the likelihood as locally Gaussian around the fiducial point, their uncertainty estimates should not be interpreted as guaranteed lower bounds on those from a full posterior analysis.

\section{MCMC likelihood}
\label{app:mcmc_likelihood}
Defining the total model spectrum in a given TDI channel as $\Omega_{\rm tot}(f_k;\mathbf{s},\mathbf{n},\mathbf{c}) \equiv \Omega_{\rm noise}(f_k;\mathbf{n}) + \Omega_{\rm foreg.}(f_k;\mathbf{c}) + \Omega_{\rm GW}(f_k;\mathbf{s})$, the combined likelihood reads
\begin{equation}
\label{eq:loglike_mixed}
\log \mathcal{L}_{\rm G+LN}(\mathbf{s},\mathbf{n},\mathbf{c}) = \frac{1}{3}\log \mathcal{L}_{\rm G}(\mathbf{s},\mathbf{n},\mathbf{c}) + \frac{2}{3}\log \mathcal{L}_{\rm LN}(\mathbf{s},\mathbf{n},\mathbf{c}) \, ,
\end{equation}
with Gaussian contribution
\begin{equation}
\label{eq:loglike_gauss}
\begin{aligned}
&\log \mathcal{L}_{\rm G}(\mathbf{s},\mathbf{n},\mathbf{c}) \\
&\quad = -\frac{N_c}{2}\sum_k n_k \left(\frac{\bar D_k - \Omega_{\rm tot}(f_k;\mathbf{s},\mathbf{n},\mathbf{c})}{\Omega_{\rm tot}(f_k;\mathbf{s},\mathbf{n},\mathbf{c})}\right)^{\!2} ,
\end{aligned}
\end{equation}
and log-normal contribution
\begin{equation}
\label{eq:loglike_lognormal}
\log \mathcal{L}_{\rm LN}(\mathbf{s},\mathbf{n},\mathbf{c}) = -\frac{N_c}{2}\sum_k n_k \log^2 \frac{\Omega_{\rm tot}(f_k;\mathbf{s},\mathbf{n},\mathbf{c})}{\bar D_k} \, ,
\end{equation}
where $\mathbf{s}$, $\mathbf{n}$ denote, respectively, the signal (cosmic-string and foreground) and instrumental-noise parameters, $\bar D_k$ is the coarse-grained data in a given TDI channel, and $n_k>1$ is the number of fine frequency values contained within the $k$-th coarse bin (with $n_k=1$ outside the coarse-grained region). The log-normal term accounts for the residual non-Gaussianity of the chunk-averaged spectral estimator, while the Gaussian term recovers the high-SNR limit, with the $1/3$--$2/3$ weighting calibrated in Ref.~\cite{Dimitriou:2025bvq}. The total log-likelihood is obtained by summing Eq.~\eqref{eq:loglike_mixed} independently over the AA, EE, and TT TDI channels,
\begin{equation}
\log\mathcal{L}_{\rm total} = \sum_{X\in\{AA,EE,TT\}} \log\mathcal{L}_{\rm G+LN}^{(X)}(\mathbf{s},\mathbf{n},\mathbf{c}) \, .
\end{equation}

MCMC sampling of the resulting posterior is performed with \texttt{emcee}, using an ensemble of $\max(64,4\,N_{\rm dim})$ walkers initialized in a small Gaussian ball around a starting point deliberately offset from the injected $\log_{10}G\mu$, in order to test recovery of the truth from a biased starting position. Convergence is monitored via the integrated autocorrelation time $\tau$, requiring both that the chain length exceed $50\,\tau$ in every dimension and that successive estimates of $\tau$, evaluated every 500 steps, agree to within $5\%$; the first $5\,\tau$ steps are discarded as burn-in. Gaussian priors on the foreground amplitudes and uniform priors on $\log_{10}G\mu$, $A_{\rm acc}$, and $A_p$ are applied identically to the priors used to train the SBI posterior (Sect.~\ref{sec:results}).

\section{SBI consistency checks}
\label{app:coverage}
\begin{figure*}[t]
\centering
\includegraphics[width=0.8\textwidth]{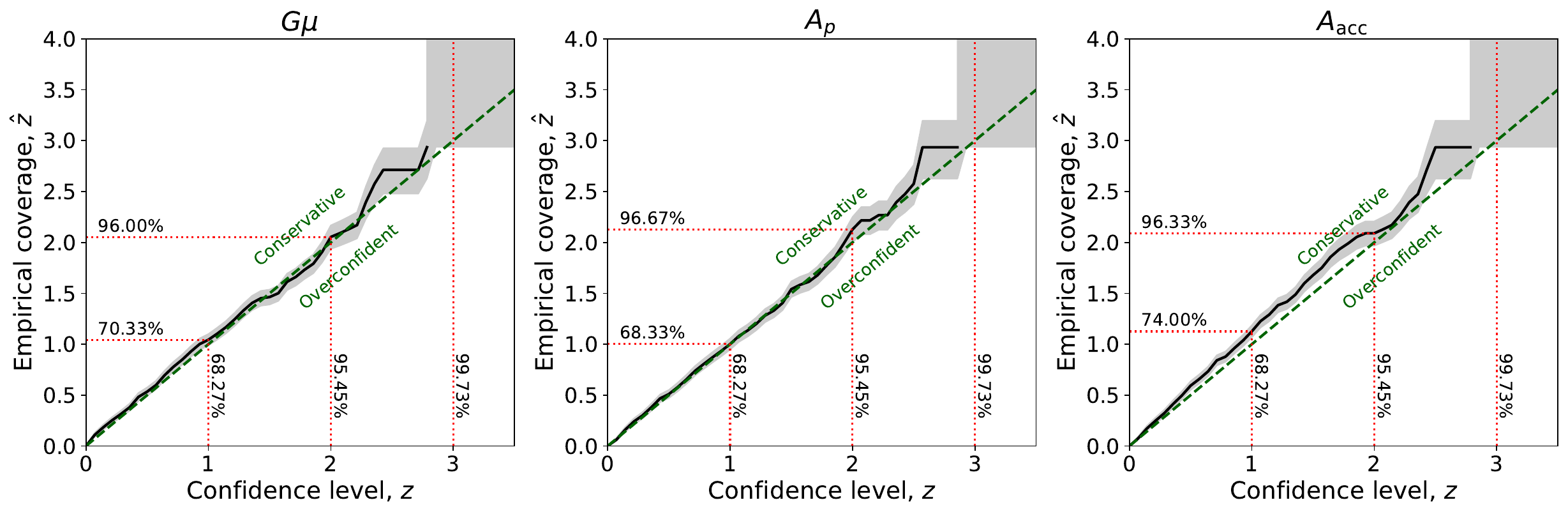}
\caption{\label{fig:coverage_nofg}
P--P plots for the foreground-free SBI model, showing empirical coverage as a function of confidence level for the three inferred parameters ($G\mu$, $A_p$, and $A_{\rm acc}$). The black curves show the empirical coverage, the green dashed lines correspond to perfect calibration, and the grey bands indicate the uncertainty due to the finite test set. The close agreement with the diagonal confirms well-calibrated posteriors.}
\end{figure*}

\begin{figure*}[t]
\centering
\includegraphics[width=\textwidth]{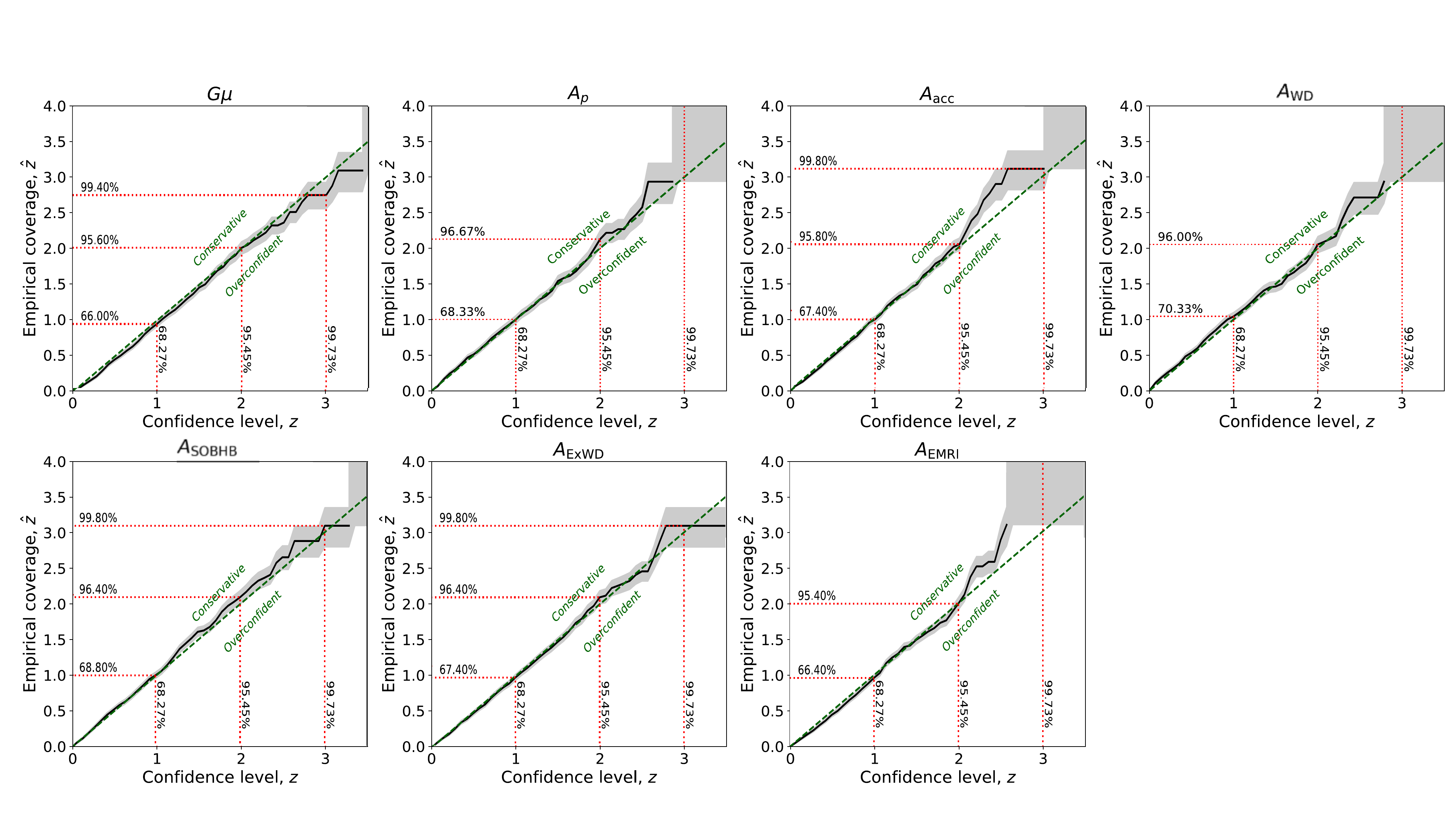}
\caption{\label{fig:coverage_fg}
P--P plots for the foreground-aware SBI model, showing empirical coverage for all seven inferred parameters: $G\mu$, $A_p$, $A_{\rm acc}$, $A_{\rm Gal}$, $A_{\rm ExB}$, $A_{\rm ExWD}$, and $A_{\rm EMRI}$. All parameters closely follow the diagonal within the finite-sample uncertainty band, demonstrating that the inclusion of astrophysical foreground nuisance parameters does not degrade posterior calibration.}
\end{figure*}

A well-calibrated posterior should reproduce the diagonal relation between empirical and nominal coverage. We demonstrate calibration in this appendix, and show resulting coverage, or P--P, 
plots in Figs.~\ref{fig:coverage_nofg} and~\ref{fig:coverage_fg}. For the foreground-free model, the empirical coverage closely follows the diagonal for all three inferred parameters ($G\mu$, $A_p$, and $A_{\rm acc}$), with empirical coverages of $70\%$, $68\%$, and $74\%$ at the $1\sigma$ credibility level and approximately $96\%$ at $2\sigma$. The foreground-aware model exhibits similarly good calibration across all seven inferred parameters, including the four astrophysical foreground amplitudes, with empirical coverage consistently close to $96\%$ at the $2\sigma$ level. In both cases the empirical coverage curves remain within the finite-sample uncertainty band over the full range of confidence levels, indicating no statistically significant evidence of over- or under-confident posterior estimates. 

\FloatBarrier

\bibliographystyle{JHEP}
\bibliography{auto,manual}

\end{document}